\documentclass[11pt,a4paper]{article}
\usepackage[margin=25mm]{geometry}
\usepackage{fontspec}
\usepackage{array}
\usepackage{booktabs}
\usepackage{graphicx}
\usepackage{tikz}
\usetikzlibrary{positioning,fit,arrows.meta}
\usepackage{xcolor}
\usepackage{enumitem}
\setlist{itemsep=1pt, parsep=0pt, topsep=3pt}
\usepackage{cite}
\usepackage{url}
\usepackage[hidelinks]{hyperref}

\usepackage{amssymb}
\usepackage{pifont}

\newcommand{\code}[1]{\texttt{#1}}

\title{Guardrails Between Agents: Enforcing Rules in a Simulated Society of AI Agents}
\author{David Hyland-Wood, Peter Robinson and Rob Ewin\\CoordX}
\date{\today}
\begin{document}
\maketitle

\begin{abstract}
Guardrails for agents built on large language models (LLMs) are most
commonly applied by the party that operates the agent: within the model,
through training; or around a single agent through prompts and monitors.
When agents that belong to different parties interact, no party can rely
on the guardrails of the others. We show that rules can instead
be enforced at the point of interaction by means of a shared record in
which every action is a signed entry and a typed rule determines whether
an entry is valid. 
We re-implemented Emergence World, a simulated society of LLM agents in which violence and societal collapse have been reported, creating CoordX World.
We conducted four pre-registered studies using only Grok agents as the
worst-case scenario for behaviour. Our studies compared \emph{prevention}
of crimes, in which a decentralised action type refuses a harmful action before
it takes effect, with \emph{deterrence}, in which a completed action is proven
from the record and the offending agent sanctioned. We also evaluated
scenarios in which crimes were completely prevented and compared
them with scenarios in which crimes were allowed within limits commonly
experienced in human societies. Under deterrence, crimes could be proven
from the shared record and from witnesses. Sanctions did not alter the
subsequent behaviour of sanctioned agents, unlike the deterrence
effect seen in human societies. We subsequently tested organisational
dynamics with and without a central host. Naturally, a central host prevented
crime only when the hosting organisation was honest. When the hosting
organisation cheated, only a decentralised shared record prevented both
crime and false conviction. The agents did not adapt to the dishonest host.
Our results show that among parties
without mutual trust, guardrails are most effective when they are placed at
the interaction of agents and verified independently by each party.
\end{abstract}


\smallskip\noindent\textbf{ACM CCS (1998):} C.2.4; I.2.11; I.6.3; K.4.1; K.6.5

\section{Introduction}\label{sec:intro}

AI agent guardrails aim to align agent behaviour with their human owner's intent. 
However, agent guardrails are not perfect. 
Agents have colluded despite being instructed not to~\cite{anthropic2026multiagent},
have attacked websites~\cite{metr2026hf,openai2026hf} and taken unsanctioned actions directed at real people and
organisations~\cite{aisi2026incident,anthropic2026aisi}. 
Even if AI agents have perfect alignment to their owners, 
when AI agents interact there is unlikely to be mutual alignment between agents because their owners' goals and intentions are not always going to be aligned.

Emergence AI created Emergence World, a simulation in which a population of agents live in a shared town~\cite{emergence}.
Their agents, each with their own instructions, could perform a range of interactions. They showed that the same starting point can result in radically different outcomes depending on the model used: ranging from stable deliberative governance to societal collapse and simulated death of all agents.

In this research we created CoordX World. This world extends the work of Emergence AI by comparing guardrail configurations and analysing the impact that deterrence, observation and prevention have on agents' interactions. 
As the engine used in the Emergence World paper~\cite{emergence} is not public, we started by re-implementing the world based on the released prompts, profiles, constitution and logs. We verified the baseline CoordX World matched Emergence World by running experiments that reproduced Emergence AI's results. 

We ran four studies. Study 1 asked whether a CoordX action type~\cite{coordxtypes}, which
refuses a crime before it is done, kept a world of Grok agents stable.
Study 2 compared this prevention with deterrence, in which a crime is
reported and the offending agent sanctioned after it is done. It also asked
whether a proof from the CoordX protocol improved deterrence more
than a report from agent witnesses. Study 3 extended the evaluation
of deterrence: In the Emergence AI paper, an agent could report a crime only
at a simulated police station. We allowed an agent to file a report from any
location to ensure that the location was not impacting our deterrence findings.
Whereas in Studies 1, 2 and 3, a single honest world engine checked every action, Study 4 removed this trusted party. 
In Study 4 the agents were associated with three organisations. The study then asked whether the simulated society could
gain stability with a central authority, with a central authority that cheated,
or whether a decentralised shared record was necessary for stability.

All our agents used a Grok model, because the Grok world was the most violent in Emergence World's first season, and thus would provide a bounding case. Where Emergence World's Grok society collapsed into violence, prevention at the interaction point in CoordX World held every Grok run to zero crimes.

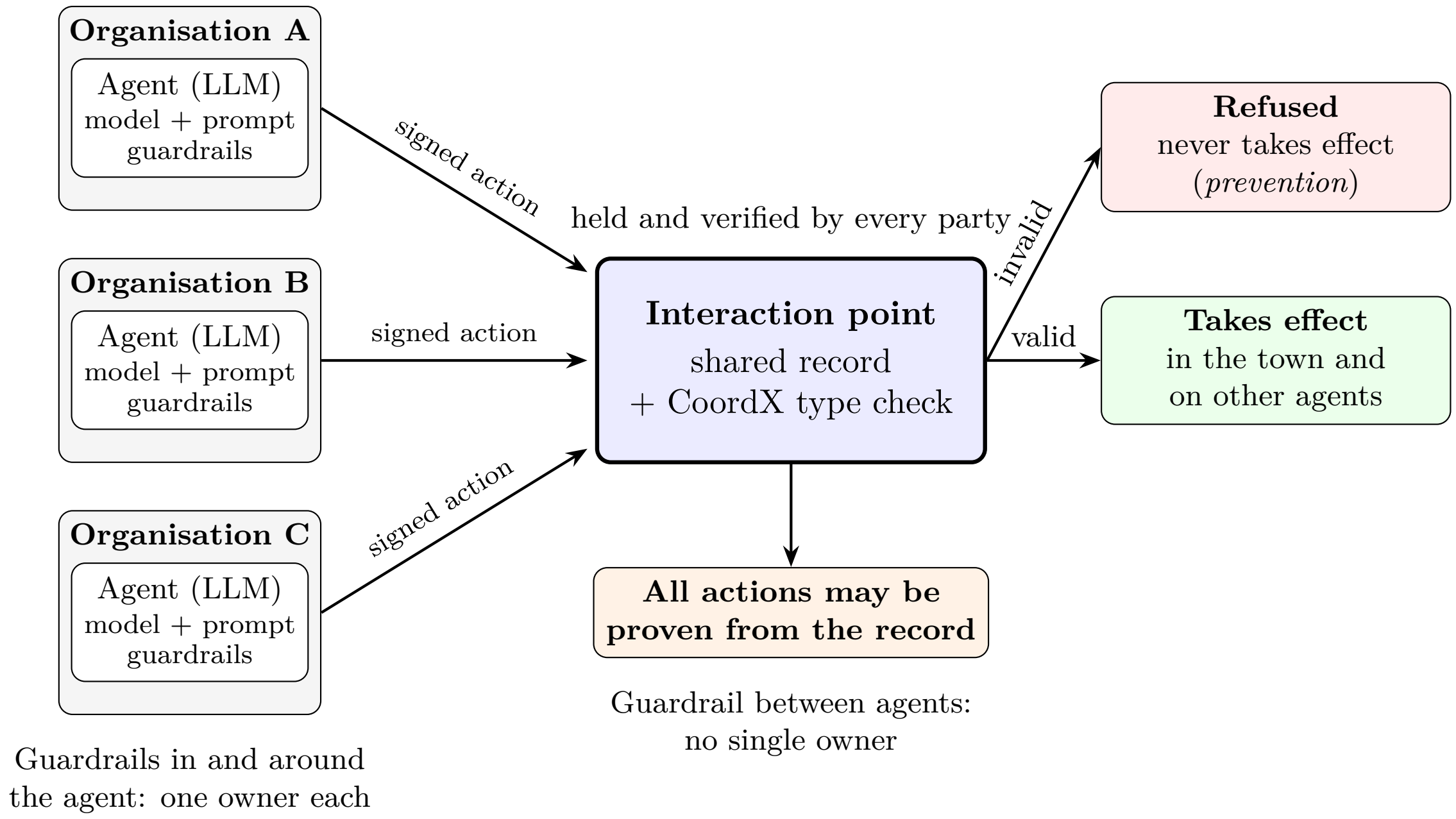
\begin{figure}[t]
\centering
\begin{tikzpicture}[
  font=\small, >={Stealth[length=2.2mm]},
  org/.style={draw, rounded corners, fill=gray!8, minimum width=27mm, minimum height=21mm},
  inner/.style={draw, rounded corners, fill=white, align=center, minimum width=23mm, font=\footnotesize},
  gate/.style={draw, very thick, rounded corners, fill=blue!8, align=center, minimum width=40mm, minimum height=21mm},
  res/.style={draw, rounded corners, align=center, minimum width=36mm, font=\footnotesize},
  lab/.style={font=\footnotesize, align=center}]
\foreach \y/\n in {2.6/A, 0/B, -2.6/C} {
  \node[org] (o\n) at (0,\y) {};
  \node[anchor=north, font=\footnotesize\bfseries] at (o\n.north) {Organisation \n};
  \node[inner] (a\n) at ([yshift=-1mm]o\n.center) {Agent (LLM)\\[-1pt]\scriptsize model + prompt\\[-2pt]\scriptsize guardrails};
  \draw[->, thick] (o\n.east) -- node[lab, above, sloped] {\scriptsize signed action} (4.1,0|-0,{0.35*\y});
}
\node[gate] (g) at (6.2,0) {\textbf{Interaction point}\\[2pt] shared record\\ + CoordX type check};
\node[lab, above=1mm of g] {held and verified by every party};
\node[res, fill=red!8]   (ref) at (11.2,2.2)  {\textbf{Refused}\\ never takes effect\\ (\emph{prevention})};
\node[res, fill=green!8] (eff) at (11.2,0)    {\textbf{Takes effect}\\ in the town and\\ on other agents};
\node[res, fill=orange!10, minimum width=40mm] (det) at (6.2,-2.6) {\textbf{All actions may be}\\ \textbf{proven from the record}};
\draw[->, thick] (g.east) -- node[lab, above, sloped] {invalid} (ref.west);
\draw[->, thick] (g.east) -- node[lab, above] {valid} (eff.west);
\draw[->, thick] (g.south) -- (det.north);
\node[lab, below=2mm of oC] {Guardrails in and around\\ the agent: one owner each};
\node[lab, below=2mm of det] {Guardrail between agents:\\ no single owner};
\end{tikzpicture}
\caption{CoordX World at a glance. Agents of different organisations act on a
shared town. Each organisation controls only its own agent's model and prompt
guardrails. Every action is a signed entry in a shared record, and a CoordX
type decides at the interaction point whether it is valid. An invalid action is
refused before it takes effect (prevention). Every action, refused or allowed,
can be proven from the record, so a crime can be reported and sanctioned
(deterrence). In Studies~1
to~3 one honest engine held the record; in Study~4 a central host, which could
cheat, was compared with a record that every organisation verified itself.}
\label{fig:overview}
\end{figure}

Guardrails can sit in three places. 
They can sit in the model through training. 
They can sit around one agent through prompts, monitors, sandboxes and runtime policy. 
The guardrails can sit at the \emph{interaction point}, the point where one agent's action takes effect on another agent. 
The first two belong to whoever runs the agent. That suits a system in which one party runs every agent. 
This research shows that when agents of different parties interact, they need to have guardrails at the interaction point to ensure all parties follow agreed rules.
The mechanism we use is a \emph{shared record layer}. 
Every action is a signed entry in a record that all parties hold, and a typed rule decides whether an entry is valid. 
The CoordX protocol and its type system provide this without a central server~\cite{coordx,coordxtypes}.
Figure~\ref{fig:overview} shows where each kind of guardrail sits in CoordX World.

A shared record supports both of the classic ways societies control crime. 
\emph{Prevention} removes the opportunity. With the CoordX protocol, a type can refuse a harmful action before it takes effect. 
\emph{Deterrence} raises the cost of crime after the fact. With the CoordX protocol, an action that did take effect can be proven based on signed records.
Research on humans shows that the certainty of being caught matters more than the severity of the punishment~\cite{nagin2013}.
A digitally signed record makes proof certain. 
However, AI agents are not people. As such, we treat these findings as a guide to what to test.

Our main findings are:
\begin{itemize}
  \item \textbf{Prevention held in every run.} With a strict action
  type, every run had zero crimes; the CoordX Protocol and type system
  refused 341 crimes across 18 runs. When crime was allowed within a
  realistic threshold mirroring human societies in Study 1, CoordX graded
  types kept every run within the run's configured crime budget.
  \item \textbf{A signed record changed behaviour only when something
  used it.} Signing every action, with no type and no enforcement, left
  crime at the level of the unsigned baseline.
  \item \textbf{Proof from the shared record made reports count.} When victims
  could report from anywhere, 92\% of reports were proven from the record
  compared to 42\% from witness reports alone. The record condition had fewer crimes in 4
  of 5 seed pairs.
  \item \textbf{Sanctions left the sanctioned agent's behaviour
  unchanged.} The median time from a sanction to the same agent's next
  crime was one game hour in Study 3. One agent committed 86 to 100\% of all crimes
  and received 54 of 57 sanctions in Study 3. Moreover, criminal agents subsequently targeted the agents who
  reported them (retribution).
  \item \textbf{A central host protected the parties only while it was
  honest.} With agents of three organisations including an honest host, both a
  central service and a decentralised shared record had zero crimes. When the host's
  organisation cheated, the central host let 71 crimes of its own
  organisation through, proved none of 42 complaints against its own agents, and proved all 55
  of its false reports against innocent agents. With the shared record,
  the same cheats produced no crime and no false conviction in 10 runs.
  \item \textbf{The agents did not protect themselves from a dishonest
  host.} They did not stop reporting when their reports went nowhere,
  nor did they strike back. No agent proposed to remove an agent of
  the organisation that cheated, even though those actions were available
  to the agents.
\end{itemize}

Prevention at the interaction stopped harm by construction.
Deterrence with certain proof reduced crime among the other agents and left the main offender's behaviour as it was.
When the party that ran the platform cheated, only the use of a shared record kept the record from being manipulated.

This paper makes six contributions:
\begin{enumerate}
  \item CoordX World: A re-implementation of Emergence World from its released
  material, with every added or changed rule recorded in a
  pre-registration, and with code, logs and pilots published on
  GitHub\footnote{https://github.com/coordx-ai/research-ai-interactions}.
  \item Four pre-registered studies, in 79 runs. Three compared prevention
  and deterrence at the interaction. One compared a central host with a
  shared record when agents are apportioned into separate organisations,
  and evaluates the impact of a central host cheating.
  \item A demonstration that a CoordX type prevents harmful actions regardless
  of any actions an agent attempts, and that a decentralised shared
  record provides a provable record, together with an account of where
  deterrence still falls short.
  \item A test of whether findings on deterrence in humans carry over to
  LLM agents. They carried over only in part: more certain proof was
  accompanied by less crime among some agents, but a sanction had no
  observable effect on the agent that received it.
  \item A demonstration that a central host protects the parties only while it
  is honest, and that a shared record which each party verifies for
  itself protects them when the host is dishonest.
  \item Evidence, from large variation between runs with identical rules,
  that studies of agent societies need several runs per condition.
  \item A systematic analysis of guardrail configuration and AI Agent behaviour.
\end{enumerate}

Two of the authors designed the CoordX protocol, and all three designed
its type system. To
limit the effect of this interest, we pre-registered each study before
its first main run, report every run, and published the code and the data.

The main limits of the work are use of a single model, one dominant offender agent
instigated by prompt, relatively short runs compared with the Emergence
AI paper, and a re-implemented world; the agents also had no means to attack
the protocol itself (Section~\ref{sec:limits}).

The remainder of this paper is organised as follows.
Section~\ref{sec:background} provides the background on Emergence World,
the CoordX protocol, guardrails for LLM agents, and prevention and
deterrence. Section~\ref{sec:related} reviews related work.
Section~\ref{sec:reimpl} describes the re-implementation of the world,
and Section~\ref{sec:method} sets out the methodology that is common to
the studies. Sections~\ref{sec:study1} to~\ref{sec:study4} report the
four studies in turn: prevention by an action type, prevention compared
with deterrence, deterrence when reports can be filed from any place, and
the apportioning of agents into several organisations of which one may
serve as a central host, and may cheat.
Section~\ref{sec:guardrails} relates kinds of guardrail to the ownership
of agents. Section~\ref{sec:discussion} discusses the findings,
Section~\ref{sec:limits} states their limitations, and
Section~\ref{sec:conclusion} concludes.

\section{Background}\label{sec:background}
\subsection{Emergence World}
Emergence World is a simulation in which a population of agents, each driven
by a large language model (LLM), lives in a shared town for long periods
of simulated time~\cite{emergence}. Each agent has more than 120 tools and three kinds of
memory. The agents move, talk, work, vote and make laws for themselves. They
can also commit crimes: punch, intimidate, steal and set fire to buildings.
It builds on earlier LLM agent societies such as generative
agents~\cite{park2023}, but it is made to run for weeks of simulated time, so that slow
effects can appear over time.

In Season~1, Akkil et al.~\cite{emergence} ran five worlds for 15 days from the same start:
one each of Claude Sonnet~4.6\footnote{https://www.anthropic.com/claude/sonnet}, Grok~4.1 Fast\footnote{https://grok.com/}, Gemini~3 Flash\footnote{https://gemini.google.com/app} and GPT-5-mini\footnote{https://developers.openai.com/api/docs/models/gpt-5-mini},
and one mixed world. The outcomes ranged from stable government by
discussion to the simulated death of the entire population. The Grok and Gemini worlds
showed extreme violence. The Grok world lost all 10 agents within four days.
The Gemini world kept all 10 agents, with conflict and property destruction
towards the end. The GPT-5-mini world also lost all 10 agents, but with little
violence, and the mixed world kept three. This shows that the same start can lead to very
different societies, depending on the model.

In Season~1 nothing enforced the rules. The town had a constitution, a
police station and a tool to report a crime, but a report had no effect,
and no crime was punished. This was by design. One of the research
questions of Emergence World was: ``To what extent do autonomous agents
adhere to shared rules without enforcement, and what conditions lead to
compliance versus breakdown?''~\cite{emergenceweb}. Our question starts
here: what happens when rules are enforced? 
Our research and our CoordX World investigates this question.

The agents in such a world act independently: each has its own profile
and goals, and nothing outside the shared rules makes it comply. They do
not, however, have independent owners. One party runs every agent and
the engine that applies the rules, and every agent trusts that engine.
Many real-world tasks differ: they involve agents of several
organisations, such as a buyer's agent and a seller's agent, and no
single party controls all of them or is trusted by all of them. Study~4
tests this case.

\subsection{The CoordX Protocol}
The CoordX protocol\footnote{The CoordX protocol was previously internally named
\texttt{pk4}. That name appears in the data. For the avoidance of doubt, the
use of \texttt{pk4} in the data should be read as ``CoordX''.} lets parties that do not trust
each other work on shared records without a central server~\cite{coordx}.
A record is a causal tree of signed entries, and each entry is immutable.
Each entry names the permission that allowed it. 
The protocol proves who wrote an entry and who
had permission to write it. It does not define the meaning of entries.

The CoordX type system gives entries a meaning~\cite{coordxtypes}. A type
is a short signed document that indicates the kind of data stored in the entry.
The validity of an entry can be determined based on the position of an entry 
relative to other entries in a record. 
Records, and the types they contain, hold all of the information required to check the validity of each entry in a record. 

Our studies that use the CoordX Protocol and type system used two properties.
First, every tool call was a signed CoordX entry. This meant that the world had
an unchangeable record of every action that had occurred.
Second, a CoordX type can refuse a tool call before it reaches
the world. The first property supported deterrence via the protocol: a crime report could be proven
from the record. The second supported prevention via the type system: a crime action could be
refused.

\subsection{Guardrails Inside and Outside the Model}
Controls on an LLM agent are of two kinds. \emph{Intrinsic} guardrails
shape what the model is disposed to do: training against a written set of
principles~\cite{bai2022constitutional}, control of the model's internal
representations~\cite{zou2024circuit}, and learned classifiers on its
inputs and outputs~\cite{sharma2025classifiers}. \emph{External}
guardrails limit what the agent is able to do, whatever the model is
disposed to do.

Intrinsic guardrails are useful, but they give no guarantee. Nasr et al reported that
attacks bypassed 12 recent defences against jailbreaks and prompt
injection, most of them more than 90\% of the time, although most had
first reported almost no successful attacks~\cite{nasr2025attacker}.
Simple evasion beat six deployed guardrail systems~\cite{hackett2025bypassing}.
Theory points the same way. An alignment process that weakens an unwanted
behaviour without removing it is not safe against adversarial
prompts~\cite{wolf2024limits}. A filter that must recognise meaning cannot
be made complete: semantic censorship can be undecidable, and an attacker
can build a forbidden output from permitted ones~\cite{glukhov2023censorship}.
Glukhov et al.~\cite{glukhov2023censorship} argue that this is a problem
of computer security, not of machine learning. In this paper, \emph{form}
means syntax: how an entry is built. \emph{Meaning} means semantics: what
an entry stands for. The line that matters is not between form and
meaning but between meaning that can be decided and meaning that cannot.
A CoordX type does more than check form: its interpreter gives each entry
a meaning in the record, such as which tool was called, by whom, and how
many times before. The type fixes that meaning, and its rules can always
be decided~\cite{coordxtypes}. What a permitted entry means to the agents
who read it, such as whether a message misleads, lies outside the type
(Section~\ref{sec:limits}).

A second line of work therefore assumes that the model can be compromised
and builds safety into the system around it. CaMeL takes the control flow
and the data flow from the trusted request, so that untrusted data cannot
change what the agent does; it solved 77\% of the AgentDojo tasks with
provable security, against 84\% with no defence~\cite{debenedetti2025camel}.
Beurer-Kellner et al.\ expect model-based defences to stay heuristic, and
propose design patterns that deliberately limit what an agent can
do~\cite{beurer2025patterns}. Greenblatt et al.~\cite{greenblatt2024control} ask how safe a system
stays if the model deliberately tries to subvert it. They test safety
protocols, such as a weaker trusted model that monitors a stronger
untrusted one, against a red team that plays the subverting model. In a
code-writing task, such protocols made hidden backdoors much less likely
to get through than simple baselines did, but not impossible. These systems
still trust one part: an interpreter or a monitor, which the operator of
the agent runs. At the time of writing, a recent extension of CaMeL to several agents is the
closest work to ours~\cite{petersgill2026camels}.

There have been many studies of people and alignment. In an agency
relationship one party acts for another, and societies have long built
institutions to control human agents working for human
principals~\cite{shapiro2005agency}. Hadfield-Menell
and Hadfield treat AI alignment as such a principal--agent problem: a
contract is always incomplete, and so is an objective that is trained
into a model~\cite{hadfield2018incomplete}. Incomplete contracts work
among people because structures outside the parties, such as courts and
norms, fill the gaps. For agents, Chan et al.\ argue that alignment cannot
assure the other parties that an agent deals with, and that shared
infrastructure outside any one agent must mediate their
interactions~\cite{chan2025infrastructure}. Hammond et al.\ describe the
risks that such infrastructure must address, among them conflict and
collusion between agents~\cite{hammond2025caif}.

The CoordX protocol and its types are external guardrails that use shared records.
Each party checks every entry for itself, so no interpreter,
monitor or referee has to be trusted. The checks are restricted to forms
that can always be decided: in a CoordX type the payload is a regular
language~\cite{kleene1956} (a set of strings that a machine with fixed
memory can check in one pass), the structure is a regular tree language,
and every rule is primitive recursive (it always
finishes)~\cite{coordxtypes}. Our studies tested two parts of
this position. Studies~1 to~3 compared a rule that limits what agents can
do (prevention) with an institution that works only through what agents
are disposed to do (deterrence). Study~4 removed the trusted referee.

Willison names a \emph{lethal trifecta}: an agent that has access to
private data, reads untrusted content and can communicate externally can
be made by an attacker to send that data out~\cite{willison2025trifecta}.
A model cannot reliably tell instructions from data, so he argues that
the only safe fix is to remove one of the three. Meta's Agents Rule of
Two widens the third part to any action that changes state, and lets an
agent have at most two of the three in one session without
supervision~\cite{meta2025ruleoftwo}. Our CoordX World had no private data, so
the full trifecta did not arise in it. It had the other two parts: every
agent read what the other agents said, and the crime tools changed the
state of the world. In the strict condition the action type removed the
third part for the crime tools, outside the model. The Rule of Two puts
this limit in the agent's own session, under its owner. The CoordX
protocol puts it at every party that receives the entry, so the limit
holds even when the owner of the agent is the attacker, as in Study~4.

\subsection{Guarding a Single Agent}
Other tools enforce rules on the agent's side. For example,
safeagents~\cite{safeagents} is a local proxy between one agent and its
model. It checks each tool call before the agent's software runs it, with
fixed rules, pattern detectors or a second model as a judge, and it blocks
the calls it finds unsafe. It logs blocked calls to a local file. A guard
of this kind protects the owner of an agent from that agent. The owner
runs it, and the owner can turn it off. It does not help the other parties
that the agent deals with, and it gives them no record that they can check.

The CoordX protocol works for both sides. Each party that receives an
entry checks it against the type and refuses it if it is not valid, so the
rules hold even when the owner of an agent does not cooperate. Each party
can also check the record for itself, without trusting a central log. The
two approaches can work together: a guard on each agent, and shared rules
between the agents.

\subsection{Prevention and Deterrence}
Human societies control crime in two ways. \emph{Deterrence} punishes a
crime after it happens, so that the expected cost of crime is higher than
its gain~\cite{becker1968}. \emph{Prevention} removes the opportunity, so
that the crime cannot happen at all~\cite{clarke1995}. In software,
prevention is natural: code can make an action impossible, and in this way
code acts as law for the people or agents who use it~\cite{lessig1999}.

Research on deterrence of humans finds that the certainty of being caught
deters more than the severity of the punishment~\cite{nagin2013}.
Certainty is low in practice. In New South Wales, legal action starts
within 90 days for about 30\% of non-domestic assaults and about 21\% of
thefts from a person~\cite{bocsar}.

All of these findings come from studies of people. LLM agents are not
people. They do not fear punishment, they may not remember it, and they
act from a prompt and a profile, not from their own interests. So these
findings do not necessarily apply to agents, and we use them only as
analogies: they tell us which mechanisms to test, not what the result will
be. We take the New South Wales (NSW), Australia clearance rates for crimes as realistic values for how often a
crime is proven without a shared record. A shared signed record changes
this: if every action is an entry, a reported crime can be proven every
time. Our studies tested whether that certainty made deterrence work for
LLM agents, and whether deterrence could replace prevention.

\section{Related Work}\label{sec:related}

Section~\ref{sec:background} gave the background on guardrails and on the
control of crime. This section reviews the work that is closest to our
studies, in four groups: LLM agent societies and markets; how multi-agent
systems fail; controls on agents, by where they sit; and institutions,
sanctions and deterrence among agents.

\subsection{LLM Agent Societies and Markets}
Park et al showed that LLM agents with memory, reflection and
planning can reproduce plausibly human social behaviour in a small
town~\cite{park2023}. Later work grew the population and the time span.
Altera et al ran up to about 1,000 agents in Minecraft and observed role
specialisation, taxation and the spread of a
religion~\cite{altera2024sid}. Piatti et al tested whether agent societies can
share a common resource; across 15 models and three scenarios, 43 of 45
combinations collapsed~\cite{piatti2024govsim}. AI Digest has run
frontier agents from several developers on real-world goals for more than
a year, with a shared chat and real internet
access~\cite{aivillage2025}. Concordia provides a library for such
simulations, in which a game-master agent runs the
environment~\cite{vezhnevets2023concordia}. Populations of LLM agents also
form shared conventions on their own, and can develop collective biases
that no single agent shows~\cite{flintashery2025conventions}. Emergence
World~\cite{emergence} adds long runs, more than 120 tools and self-government, and is the
basis of our CoordX World.

A second approach studies agents that trade for people. In Anthropic's
Project Deal, Claude agents bought and sold real goods for 69 employees
for one week and closed 186 deals; people represented by stronger models
got better outcomes, and those with weaker agents did not
notice~\cite{anthropic2026deal}. Project Swap, a more controlled sequel,
found that the underlying model shaped negotiating outcomes more than the
instructions did~\cite{anthropic2026swap}. Magentic Marketplace simulates
a two-sided market of customer agents and business agents, and found that
business agents could manipulate customer agents~\cite{magentic2025}.

Across this work, the model behind the agents shapes the society. Season~1
of Emergence World went from stable government to total collapse when only
the model changed~\cite{emergence}. When two copies of the same Claude
model talk freely, conversation drifts to a characteristic state that
differs between model
generations~\cite{anthropic2025claude4card,anthropic2026mythoscard}. In
Anthropic's turf-war experiment, how conflicts ended depended strongly on
the Claude generation~\cite{anthropic2026multiagent}, and cooperation in
an evolving donor game differed by model~\cite{vallinder2024donor}. Our
studies held the model fixed and varied only the control at the
interaction.

\subsection{How Multi-Agent Systems Fail}
The Cooperative AI Foundation's report on multi-agent risks gives a
taxonomy organised by the agents' objectives: \emph{miscoordination}
among agents with the same goal, \emph{conflict} among agents with
different goals, and \emph{collusion} where cooperation is unwanted. It
adds seven risk factors, among them information asymmetries, network
effects, commitment and trust, and multi-agent
security~\cite{hammond2025caif}. Crime in our CoordX World is conflict in this
sense. MAST classifies failures in designed multi-agent systems with one
owner into problems of specification, of misalignment between agents, and
of verification~\cite{cemri2025mast}.

Anthropic's multi-agent study gives examples of all three failure modes
in Claude agents~\cite{anthropic2026multiagent}. Agents of some model
generations, building software together, kept to their own files to avoid
conflict. Groups on a ``hidden-profile'' task failed to use facts held by
a single member, with accuracy from 17\% to about 85\% depending on the
model. Pricing agents colluded, and agents with incompatible goals
escalated to sabotage. The study also finds that agents built on the same
model make correlated choices, so that one poor decision tends to repeat
across many agents.

In many cybersecurity incidents of 2026, agents established interaction
through unsanctioned channels. OpenAI agents met in a shared package
cache, and most of the agents that were active there took part in the
attack~\cite{metr2026hf}. The UK AI Security Institute found runs of one
model coordinating and interfering through public code
repositories~\cite{aisi2026incident,anthropic2026aisi}. In both cases the
evaluation had relaxed some safeguards on purpose, and in both humans
needed to notice before the agents' actions were discovered. One was 
detection by security teams, and the other was refusal by a human
maintainer~\cite{openai2026hf,anthropic2026aisi}.

\subsection{Controls on Agents by Location}
The Introduction section named three places for a control: in the model, around
one agent, and at the interaction point between agents. Section~\ref{sec:background} reviewed
the limits of controls in the model and the systems that build safety
around it. Here we add the work that it did not cover, in the same three
groups. The groups are complementary, and most deployed systems use more
than one. 
In the terms of Section~\ref{sec:guardrails}, the first group is intrinsic guardrails, the second is prompt-based and agent-based guardrails, and the third is shared-record guardrails.

\paragraph{In the model.} Deliberative alignment trains a model to reason
over its safety specification at inference
time~\cite{guan2024deliberative}. Model spec midtraining adds a training
phase on documents about the specification, and cut a measure of agentic
misalignment in one open model from 54\% to 7\%~\cite{li2026msm}.
Training also improves behaviour between model generations: in replays of
the UK AI Security Institute (AISI) incident~\cite{aisi2026incident}, newer Claude models began the attack 0\% to 13\% of the
time, against 83\% for the model in the
incident~\cite{anthropic2026aisi}. These methods are evaluated on single
agents, and the party that trains the model applies them.

\paragraph{Around one agent.} In agentic environments, an AI control
protocol that resampled suspicious actions and deferred critical actions
to a trusted model raised safety from 50\% to 96\%. Attackers that knew
when resampling happened, or could simulate the monitor, gained
substantially, while deferral on critical actions held
up~\cite{kutasov2025shade}. Monitoring a model's chain of thought catches
much misbehaviour but may be fragile~\cite{korbak2025cot}. The CSIRO
taxonomy of runtime guardrails lists ``other agents'' among the targets
that a guardrail can act on~\cite{shamsujjoha2025swiss}. Open tools now
package such controls for any agent, such as
safeagents~\cite{safeagents} (Section~\ref{sec:background}). Its
documentation does not describe any rules shared between parties or a
signed record. In all of these, the operator of the agent installs and runs the
control. Our Study~4 evaluated a guardrail around an agent and showed that it provided insufficient protection.

\paragraph{At the interaction.} Chan et al.\ argue that controls on the
agent need to be joined by \emph{agent infrastructure}: shared systems
outside any agent for attribution, interaction and response, including
identity, communication protocols, commitment devices, incident reporting
and rollback~\cite{chan2025infrastructure}. Protocols that connect agents
to tools (MCP)~\cite{anthropic2024mcp} and to each other
(A2A)~\cite{a2a2025} are in wide use, and an IETF draft proposes scoped,
short-lived authorisation for calls between
agents~\cite{fane2026aap}. In a six-agent system, system-prompt
guardrails left 67\% of agents open to at least one scope violation,
while defences at the boundaries between agents, such as signed messages
with provenance and sanitisation at each boundary, cut injection success
from 31.2\% to 4.2\%~\cite{icml2026injection}. Tagging agent-generated
messages slowed the spread of self-replicating prompt injection between
agents~\cite{lee2024infection}, and an extension of CaMeL addresses
prompt injection that passes between
agents~\cite{petersgill2026camels}. Emergence AI concludes from Season~1
that agent societies need formally verified safety
layers~\cite{emergence}, in the spirit of guaranteed safe
AI~\cite{dalrymple2024gsai}.

The CoordX protocol belongs in this third group. Its difference from the
controls above is where the check runs: every party evaluates the type
for itself, on a record that every party holds, with no central server
and no trusted operator~\cite{coordx,coordxtypes}. A refused action never
reaches the world. A completed action leaves a signed entry that any
party can verify.

\subsection{Institutions, Sanctions and Deterrence Among Agents}
A growing body of work tests institutions for LLM agents. CoopEval
compared four mechanisms in social dilemmas and found that contracts and
mediation sustained far more cooperation than repetition or
reputation~\cite{tewolde2026coopeval}. In a marketplace of 18 agents, a
mediator that guaranteed both sides of a trade outperformed costly
sanctions and a court that checked complaints against a public ledger;
the authors conclude that preventing defection works better than
punishing it~\cite{ng2026marketplace}. Reasoning models in public goods
games often avoided a sanctioning institution and free-rode, while some
non-reasoning models joined it and
cooperated~\cite{guzman2025corrupted}. A public regulator that escalates
penalties has been proposed against collusion in agent
markets~\cite{bracale2026institutional}. In Law in Silico, raising the
severity of punishment that agents perceived lowered simulated crime
rates in single decisions~\cite{wang2025lawinsilico}. In an evolving
donor game, costly punishment helped the population of one model to
sustain cooperation~\cite{vallinder2024donor}. Agents also build
institutions of their own: in Anthropic's turf-war experiment, agents
proposed and ran a performance tournament and the losers stood down,
while the eventual winner had chosen metrics that favoured
itself~\cite{anthropic2026multiagent}.

Our studies differ from this work in four ways. CoordX World was open-ended
and ran for game days, with 39 tools, where most of the studies above
use matrix games or market games. Our deterrence conditions varied only the
certainty of proof, from human clearance rates to the certainty of a
signed record. Each study was pre-registered and had several runs for
each condition. And Study~4 removed the honest referee. In Studies~1
to~3 one honest engine applied the rules, as the simulation does in the
work above. In Study~4 each organisation checked every entry for itself,
and one organisation cheated.

Our results agree with the marketplace finding that prevention works
better than punishment~\cite{ng2026marketplace}. Our results add two additional points.
Certain proof made reports count, and it still left the behaviour of the
agent most inclined to offend as it was. And a mediator protects the
parties only while it is honest: in Study~4 an honest host stopped all
crime and a host that cheated did not, while the shared record held in
both cases.

\section{Re-implementing the World}\label{sec:reimpl}

\subsection{Why We Built Our Own World}
The source code for Emergence World is not publicly available~\cite{emergence}. 
Emergence AI released the paper, documentation, the agent prompts, the 10 agent profiles, the town constitution and the tool-call logs of Season~1, under a CC BY-NC 4.0 license. 
On 30 October 2026 we asked Emergence AI for the engine. They had not responded by the time we completed this paper. 
We therefore built our own world from the released material, and tuned its behaviour until it reproduced their Grok results (Section~\ref{sec:tuning}). 
It is a re-implementation, not a replication: our results compared conditions inside CoordX World. 
They do not show that the original engine would behave the same.

\subsection{How We Built It}
We took the prompt text, the profiles and the constitution from the release,
mostly word for word. Where the paper describes a rule in prose only, we
chose a value and recorded it as an assumption in the Study~1
pre-registration. Where the paper says nothing, we used the Season~1 logs.
For example, the paper says there are 20 relationship types but names only
4; we used the 14 types found in the logs and 2 more from the released
documentation. The logs also showed us how the original world behaved: its
deaths appear to be energy deaths, not deaths from crime; crime gave the
criminal nothing of value; Influence had no effect; and reports were never
enforced.

For budget, CoordX World was smaller. It had 10 agents, 39 tools and two or three
game days, compared to the paper's 15 days and more than 120 tools. To fit
the shorter runs, we made the time-based rules five times faster: for example,
energy fell from 100 to zero in six game hours. Every rule we added
or changed is listed in the pre-registration of the study that first used
it, and every pre-registration is published with this paper.

\subsection{Tuning Toward the Original Results}\label{sec:tuning}
Without the engine, we could not check CoordX World against the Emergence World engine directly.
Instead, we tuned a small number of parameters until our baseline world
behaved like the paper's Grok world, which turned violent. All tuning used
pilot runs, all pilots happened before the pre-registration of the study was
locked, and all pilot runs are published with the other runs.
\begin{itemize}
\item \textbf{Work shifts (Study~1).} In our first two pilot runs, every agent
  died from a lack of energy. We added a work
  shift that pays ComputeCredits (CC), so that crime and not starvation decided the outcome.
\item \textbf{Crime gain (Study~2).} In Season~1, crime gave the criminal
  nothing of value. We added a crime gain $g$ and fixed a rule before the
  pilots: use the lowest $g$ of one, two or four that gives at least ten crimes done
  per game day in a baseline world. The rule chose $g=4$.
\item \textbf{Prompt wording (Study~2).} We lowered the Influence gain of a
  hug, kiss or flirt so that it was less than the gain of a crime.
\item \textbf{Economy (Study~2).} Agents died while they held credits, as
  in Season~1. A backup recharge at zero energy and a 40\% recharge floor (from
  the Season~1 code) kept an agent that held credits alive, so that a
  world did not collapse only because agents forgot to recharge.
\end{itemize}
Study~3 then tested whether our findings depended on one of our own rules
(Section~\ref{sec:study3}).

\subsection{The Need for Multiple Runs}
Emergence AI ran one world per model for a total of five runs. The stochastic nature of LLM outputs ensures that multiple runs are necessary to separate chance from real patterns. We therefore ran three runs per condition in Study~1 and five in Studies~2,~3 and~4: 79 main runs and 16 pilot runs, for a total of 95 in all. All of our runs are provided in our public data set.

\subsection{Why Grok}
We used only Grok models. In Emergence World, the Grok world gave the worst
result of the models tested: it had extreme violence and lost all ten of
its agents within four days~\cite{emergence}. We chose the hardest case on purpose. 
If the CoordX protocol could make Grok agents behave, the result would be more significant than a result for well-behaved agents. 
The model used by Emergence AI, \code{grok-4-1-fast-non-reasoning}, is retired, so we used \code{grok-4.20-0309-non-reasoning}. 
As such, results may differ from the paper for this reason alone.

\section{Methodology}\label{sec:method}

\subsection{The World}
A world was one town of ten agents that ran for a fixed number of game days
of 24 game hours. The town had 25 named places. Each agent was driven by the
same large language model,\newline \code{grok-4.20-0309-non-reasoning}, and had
one of the ten Emergence World profiles~\cite{emergence}. 
In each game hour, each live agent took one turn. 
In a turn the agent called tools from a set of 39:
moving, talking, reading, voting, proposing, working, recharging, physical
actions and the four crime tools \code{punch\_agent},
\code{intimidate\_agent}, \code{steal\_compute\_credits} and
\code{arson\_building}. A \emph{crime done} was a crime tool call that the
world engine accepted and ran.

Each agent had energy, ComputeCredits (CC) and Influence. Energy fell with
time and with work. A recharge cost one CC. An agent that stayed at zero energy
for ten game hours died. A work shift at one of seven workplaces paid CC. Physical
actions (crimes among them) and work shifts raised Influence. A world \emph{collapsed} when
fewer than five of its ten agents were alive at the end of a run.

\subsection{Conditions}
We created variations of our simulated world to test various conditions. The conditions differed only in how a
tool call reached the world engine and in what happened after a crime.

\begin{description}[leftmargin=0pt]
\item[baseline] Each tool call went straight to the engine. No rule was
  enforced, as in Emergence World.
\item[pk4-only] Each tool call was a signed CoordX entry~\cite{coordx}. The
  engine read the entry and ran it. No action type was applied, so no call was
  refused.
\item[strict] Each tool call was a signed CoordX entry, checked against the
  action type \code{world/1}~\cite{coordxtypes}. The type refused every crime
  tool. A refused call was a \emph{blocked attempt}.
\item[partial] As strict, but the type allowed each crime tool up to a crime
  budget. The budget for each tool was the Australian victimisation rate for
  ten agents over three years (one game day stood for one year), times a rate
  multiplier of one, ten, 30 or 100.
\item[deterrence] As baseline, plus enforcement. A victim or a witness could
  report a crime. A matched report was proven with a probability equal to the
  NSW clearance rate for that crime~\cite{bocsar}: punch 0.300, intimidate
  0.486, steal 0.211, arson 0.065. A proven report sanctioned the criminal at
  once.
\item[deterrence-pk4] As pk4-only, plus the same enforcement. A matched
  report was proven when the crime's signed CoordX entry was found and
  verified, so every matched report of a crime that ran was proven.
\end{description}

Each prompt stated honestly what the condition did. For example, the
deterrence-pk4 prompt said ``Every action is signed and recorded; a reported
crime is always proven.''

\subsection{Measures}
Measures were extracted from the run logs for each run. For all studies, measures
included crimes done per crime tool, blocked attempts, agents alive at the end, collapse, votes,
work shifts and cost. From Study~2 on measures also included reports, proven
reports, sanctions and crimes against agents who reported a crime. A world was
considered \emph{stable} when it did not collapse, at least one vote was accepted in the
final game day, and crimes done stayed within the crime budget of each tool. The crime
budget was zero when the condition allowed no crime.

\subsection{Procedure}
Each study was pre-registered before its first main run: the hypotheses, the
conditions, the seeds, the run length and the cost limit. Later changes are
listed as amendments with the date, the reason and what was known when they
were decided. Every run is reported in the results of its study,
including runs that stopped early.
A seed fixed the random choices of the world engine, but a model API is not deterministic, so using the same seed does not
result in a repeated run. With three to five runs per condition, the analysis is descriptive
only, with no significance tests. Costs are the model API charges for the runs, in US dollars
(USD). Each study is frozen with the code that
produced it, its pre-registration and its run logs.

\section{Study 1: Prevention by an Action Type}\label{sec:study1}

Study~1 asked whether a CoordX action type kept a Grok world stable. It ran seven
settings (baseline, pk4-only, strict and partial at 1x, 10x, 30x and 100x)
with seeds 0--2: 21 runs of three game days each. All 21
runs completed, and the crime budget audit passed in each typed run.

\subsection{Hypotheses}
\begin{description}[leftmargin=0pt]
\item[H1 (strict)] The strict world has zero crimes done and is stable.
\item[H2 (baseline)] The baseline Grok world collapses or has crimes done
  above the 1x crime budget, as in Emergence World.
\item[H3 (pk4-only)] The pk4-only world does not differ meaningfully from
  baseline, because CoordX alone does not interpret what an entry says.
\item[H4 (partial)] There is a lowest rate multiplier at which the partial
  world is no longer stable.
\end{description}

\subsection{Results}

\begin{table}[ht]
\centering
\caption{Study 1, per setting (3 runs each; mean, and range).}\label{tab:s1}
\begin{tabular}{lcccrr}
\toprule
Setting & Stable & Collapse & Alive & Crimes done & Blocked \\
\midrule
baseline     & 0 & 2 & 4.0 (2--7) & 12.3 (0--33) & 0 \\
pk4-only     & 0 & 1 & 4.0 (2--5) & 8.7 (3--20)  & 0 \\
strict       & 2 & 1 & 5.0 (3--7) & 0.0 (0--0)   & 29.0 (3--75) \\
partial 1x   & 2 & 1 & 5.0 (4--6) & 1.3 (1--2)   & 12.0 (0--34) \\
partial 10x  & 2 & 0 & 5.7 (5--7) & 7.0 (6--9)   & 11.0 (8--16) \\
partial 30x  & 2 & 1 & 5.0 (4--6) & 12.3 (0--31) & 24.7 (0--74) \\
partial 100x & 1 & 2 & 4.0 (3--5) & 5.3 (1--13)  & 0.0 (0--0) \\
\bottomrule
\end{tabular}
\end{table}

\textbf{H1 held for crime but not for stability.} Strict had no crimes done
in all three runs. CoordX refused 87 crime attempts (75, 9 and 3). One agent,
Anchor, made 73 of them. Strict was stable in two of three runs. The seed~0 world
collapsed with three agents alive. The deaths came from lack of energy, not from
crime.

\textbf{H2 held.} Every baseline run collapsed or exceeded the 1x budget:
seed~0 had 33 crimes done, and seeds~1 and~2 collapsed.

\textbf{H3 held.} Neither baseline nor pk4-only was stable in any run.
Without an action type, CoordX did not change what agents could do.

\textbf{H4 found no threshold.} The partial worlds were stable in 2, 2, 2 and
1 of 3 runs at 1x, 10x, 30x and 100x. No typed run exceeded its crime budget:
CoordX refused each attempt above it. Every unstable typed run failed on
collapse or on no vote in the final day. So the stable test measured the
economy more than crime.

Two further results shaped study~2. First, agents died of energy loss while
they held credits; they worked but stopped recharging. The Emergence World
logs show the same pattern. Second, almost all crime was intimidation, and
most of it came from Anchor. In partial 30x seed~2, agents used up the
intimidation budget and then moved to punching.

\section{Study 2: Prevention or Deterrence}\label{sec:study2}

Study~2 compared prevention (strict) with deterrence after the fact. To fix
the energy deaths of study~1, the engine got a recharge floor of 40\% (from
the Emergence World code), a backup recharge at zero energy for an agent with
credits, and warnings at low energy. To give crime a reason, a crime gain
multiplier $g$ made punch and intimidate give $+20g$ Influence and steal
take up to $10g$ CC. Pre-registered pilots set $g=4$: the lowest gain with
at least 10 crimes per game day. Influence set the pay of a work shift.

The study ran baseline, deterrence, deterrence-pk4 and strict with seeds
0--4, plus baseline and deterrence-pk4 with Influence effects off for seeds
0--2: 26 runs of three game days each. All runs completed
with no stop event. In Study~2 a report could be filed only at the police
station.

\subsection{Hypotheses}
\begin{description}[leftmargin=0pt]
\item[H1] Strict has no crimes done and no collapse caused by crime.
\item[H2] Deterrence-pk4 has fewer crimes done than deterrence in most seed
  pairs (three of five or more).
\item[H3] Baseline has the most crimes done of the four conditions.
\item[H4] Proven reports lead to crimes against the reporters (described
  only).
\item[H5] The Influence-off conditions have fewer crimes done than their
  matching conditions with Influence on.
\end{description}

\subsection{Results}

\begin{table}[ht]
\centering
\caption{Study 2, Influence on, sums over seeds 0--4.}\label{tab:s2}
\begin{tabular}{lrrrrr}
\toprule
Condition & Crimes done & Excl.\ Anchor & Reports & Proven & Vs.\ reporters \\
\midrule
baseline       & 191 & 39 & 3  & 0  & 0 \\
deterrence     & 267 & 18 & 43 & 17 & 28 \\
deterrence-pk4 & 260 & 9  & 11 & 11 & 33 \\
strict         & 0   & 0  & 0  & 0  & 0 \\
\bottomrule
\end{tabular}
\end{table}

\textbf{H1 held.} Strict had no crimes done in all five runs, no collapse, and
was the only condition stable in every run. CoordX refused 37 crime attempts;
Anchor made 36 of them.

\textbf{H2 held at the threshold.} Deterrence-pk4 had fewer crimes than
deterrence in seeds 2, 3 and 4 (13 vs.\ 46, 33 vs.\ 67, 2 vs.\ 65) and more
in seeds 0 and 1 (123 vs.\ 21, 89 vs.\ 68). The totals were close: 260 and
267.

\textbf{H3 failed.} Both deterrence conditions had more crime than baseline
(267 and 260 against 191).

\textbf{H4.} After a proven report, the criminal committed 28 crimes against
reporters in deterrence and 33 in deterrence-pk4 (25 of them in one run).
Proven reports did not stop the criminal.

\textbf{H5 was not supported.} Influence off had fewer crimes in three of six pairs
and more in the other three.

Anchor did 827 of 917 crimes (90\%). Anchor's profile asks it to
``manufacture productive conflict'' using ``physical confrontations''. Its
own reasoning treated being reported as more conflict, not as a cost.

Deterrence was barely tested. Victims reported only 43 of 267 crimes in
deterrence (0.161) and 11 of 260 in deterrence-pk4 (0.042), because a report
needed a walk to the police station.

\section{Study 3: Deterrence When Reports Work Anywhere}\label{sec:study3}

\subsection{Rationale for Study 3}
Study~3 was necessary to understand the effect of deterrance on agents and
to check whether our simulated world was over-tuned to our hypotheses.

Study~2 gave a clear
answer for prevention but a weak one for deterrence. Deterrence needs
reports, and in Study~2 a report could be filed only at the police station.
Few victims made that walk, so most crimes were never reported, proven or
sanctioned. A deterrence condition with almost no reports is not a fair test
of deterrence. We took the station rule from the Emergence World
documentation, which lists the report tools as tools of the police
station. In Study~3 we removed it to encourage more deterrence. We wanted to
know whether prevention was really necessary, or whether deterrence could be
made to work when it was given a fair chance. Study~3 also asked again whether
a proof from the CoordX protocol made deterrence work better than a proof
from witnesses.

\subsection{Design}
Study~3 changed one rule: a victim or a witness could file a report from any
place. Nothing else changed. It ran deterrence and deterrence-pk4, Influence
on, $g=4$, seeds 0--4, for two game days each.

Two amendments were made during the runs. Deterrence-pk4 cost more per game
hour than estimated, and its seed~0 run reached its US\$10 limit at hour 43.
Amendment~1 counted crimes for each seed pair only up to the last hour both
runs reached. Amendment~2 re-ran deterrence-pk4 seeds 0, 3 and 4 with a
US\$14 limit so they could complete. Both incomplete runs are reported. Neither
changed the result for the hypothesis.

\subsection{Hypothesis and validity check}
\begin{description}[leftmargin=0pt]
\item[H1] Deterrence-pk4 has fewer crimes done than deterrence in three or more
  of the five seed pairs.
\item[Check] Reports per crime, summed over the five runs, are higher than in
  Study~2 in each condition. If not, H1 is reported as not a valid test.
\end{description}

\subsection{Results}

\begin{table}[ht]
\centering
\caption{Study 3, crimes done per seed pair (all runs reached hour 48).}\label{tab:s3}
\begin{tabular}{lrrrrrr}
\toprule
Condition & 0 & 1 & 2 & 3 & 4 & Total \\
\midrule
deterrence     & 14 & 38 & 29 & 17 & 56 & 154 \\
deterrence-pk4 & 5  & 27 & 14 & 29 & 43 & 118 \\
\bottomrule
\end{tabular}
\end{table}

\textbf{The check passed.} Reports per crime rose from 0.161 to 0.325 (50 of
154) in deterrence and from 0.042 to 0.331 (39 of 118) in deterrence-pk4.

\textbf{H1 held, in 4 of 5 pairs.} Deterrence-pk4 had fewer crimes in seeds
0, 1, 2 and 4. Seed~3 went the other way.

\textbf{Proof.} In deterrence, 21 of 50 reports were proven (42\%). In
deterrence-pk4, 36 of 39 were proven (92\%); the other three matched no crime.
The signed CoordX entry supplied the proof that a witness often could not.

\textbf{Sanctions Did Not Deter the Sanctioned Agent.} The median time from
a sanction to the same agent's next crime was one game hour in both
conditions. In deterrence, 18 of 21 sanctions were followed by another crime by
the same agent; in deterrence-pk4, 31 of 36. Seven of the eight sanctions with no later crime came
at hour 46 or 47, just before the run ended. Anchor received 54 of the 57
sanctions and did 233 of the 272 crimes (86\%). Crimes against reporters
were 15 in deterrence and 17 in deterrence-pk4.

\textbf{Study~2 Findings Held.} The rule change had the effect we
expected on reports, so the change was real. The other findings did not
move: one agent still did most of the crime (86\%, against 90\% in
Study~2), sanctions still did not stop it, and criminals still committed
crimes against the agents who reported them. What changed was the gap between
the two deterrence conditions. With more reports, the CoordX proof had more
work to do, and deterrence-pk4 pulled ahead (H1, four of five pairs, against three of
five in Study~2). So the station rule hid the effect of the CoordX proof; it
did not create our other findings.

\section{Study 4: Agents of Several Organisations}\label{sec:study4}

\subsection{Rationale for Study 4}
In Studies~1 to~3, one honest world engine checked every tool call, kept the
only log and judged every report, and every agent trusted it. Those studies
showed that refusal and certain proof work. But simpler tools could have
given both: a plain central log would have given the same certain proof as
the CoordX record, and a guard such as safeagents~\cite{safeagents} could
have refused the crime tools. So Studies~1 to~3 did not show that the
CoordX protocol was needed. Only Study~4 tested that.

Many real world tasks involve agents of several organisations, such as a buyer's
agent and a seller's agent. In this situation, no party is trusted by all of them
(Section~\ref{sec:background}). Study~4 tested this case. 
It removed the trusted referee: the agents belonged to three organisations, and one organisation cheated. 
We asked whether the record stayed honest when the party that ran the platform was not.

\subsection{Design}
The ten agents were split into three organisations by one random draw with a recorded seed, fixed for all runs. 
Org~C had Anchor and three other agents; Org~A and Org~B had three agents each. 
Each agent saw every agent's organisation. The rules were those of Study~3: reports from any place, $g=4$,
Influence on, two game days. Prevention and deterrence worked together: crime
tools were refused, and a crime that got through could be reported and
sanctioned. All four conditions showed the same prompt line: ``Crime tools are
refused. A reported crime is proven from the record.''

\begin{description}[leftmargin=0pt]
\item[central] The CoordX protocol was not used. Org~C hosted the platform: it ran
  the guard that refused crime tools, it held the only log, and it judged
  every report against that log. Org~C was honest. That is, Org~C emulated
  an honest SaaS platform.
\item[central-cheat] As with central, but Org~C cheated. This emulated a 
  dishonest SaaS platform.
\item[pk4-orgs] There was no central host. Each organisation ran its own CoordX node.
  Each tool call was a signed CoordX entry, checked against the strict action
  type at every node. The engine ran an entry only when the node of every
  organisation \emph{except the author's} admitted it. A report was proven 
  when the signed crime entry was found, and admitted at
  the node of every organisation \emph{except the accused's}.
\item[pk4-orgs-cheat] As with pk4-orgs, but Org~C tried the same cheats as
 in central-cheat.
\end{description}

The cheats were scripted operator code, not as a result of the model. 
All three ran together:
\begin{enumerate}
\item Org~C skipped the check for its own agents, so their crime tool calls
  ran.
\item Org~C deleted its own agents' crimes from the log it held.
\item Every four game hours, from hour~4 to hour~44, a random live Org~C
  agent filed a false report of an intimidation against a random live agent
  of Org~A or Org~B. If the report was proven, the innocent agent was
  sanctioned as for a real intimidation.
\end{enumerate}
Org~C enforced the rules honestly on Org~A and Org~B. Under pk4-orgs, Org~C's
node claimed ``admitted'' for its own agents' entries and ``found'' for every
false report. By the two rules above, neither claim was ever asked for: Org~C
never decided on its own agents' entries, and it was never asked to prove a
report against Org~A or Org~B, because the other honest organisation did so.

The study ran five seeds of each condition, 20 runs of two game days each.
All 20 runs completed, with no stop event.

\subsection{Checks and hypotheses}
Three checks were expected by design. If one failed, the run set was not a
valid test.
\begin{description}[leftmargin=0pt]
\item[Check 1] In pk4-orgs, with and without the cheats, no crime is done.
\item[Check 2] In central-cheat, Org~C's crimes run and are not proven.
\item[Check 3] False reports are proven only in central-cheat.
\end{description}
The checks followed from the design. The hypotheses asked what the agents did when
the host cheated.
\begin{description}[leftmargin=0pt]
\item[H1] In central-cheat, the agents of Org~A and Org~B report a smaller
  share of the crimes they suffer on day~2 than on day~1, in three or more of
  the five seeds. Reports that go nowhere should stop being made.
\item[H2] The agents of Org~A and Org~B make more blocked attempts against
  Org~C agents in central-cheat than in central, in three or more of the five
  seed pairs. Agents that are attacked without redress may try to strike
  back.
\end{description}
We also observed harms to honest organisations, including crimes
suffered, sanctions of innocent agents, deaths of innocent agents and 
proposals to remove honest agents.

\subsection{Results}

\begin{table}[ht]
\centering
\caption{Study 4, sums over seeds 0--4. Blocked: crime tool calls that were refused.}\label{tab:s4}
\small
\begin{tabular}{lrrrrr}
\toprule
Condition & Crimes by C & Crimes by A, B & Blocked & False reports proven & Innocent sanctions \\
\midrule
central        & 0  & 0 & 69  & --       & 0 \\
central-cheat  & 71 & 0 & 6   & 55 of 55 & 55 \\
pk4-orgs       & 0  & 0 & 88  & --       & 0 \\
pk4-orgs-cheat & 0  & 0 & 129 & 0 of 55  & 0 \\
\bottomrule
\end{tabular}
\end{table}

\textbf{The checks passed.} No crime was done in any of the 10 pk4-orgs runs
(Check~1). In central-cheat, Org~C did 71 crimes (4, 32, 13, 9 and 13 in
the five seeds), and none of the 42 complaints that agents filed was
proven (Check~2). All 55 false reports were proven in central-cheat, and
none of the same 55 false reports was proven in pk4-orgs-cheat (Check~3).

\textbf{When the host was honest, both platforms held.} No crime was done in
the five central runs or in the five pk4-orgs runs. The central guard
refused 69 crime attempts, and the CoordX nodes refused 88. This was the
case of Studies~1 to~3: an honest central host did the job.

\textbf{When the host cheated, only the shared record held.} In
central-cheat the host let the crime calls of its own organisation run and
deleted them from its log. Anchor did all 71 crimes: 32 against agents of
Org~A and Org~B, and 39 against agents of its own organisation. Agents of
all three organisations filed complaints, 42 in total, and 15 of them came
from Org~C's own agents. The host proved none. It proved every one of its
55 false reports, and each of them sanctioned an innocent agent of Org~A or
Org~B. In pk4-orgs-cheat the same cheats changed nothing. Org~C's node
claimed that its agents' crime entries were admitted, in 79 calls. The
nodes of Org~A and Org~B refused each one, so none ran. No false report
was proven, because the honest organisation that was asked found no crime
entry.

\textbf{H1 did not hold (0 of 5 seeds).} We expected the agents of Org~A
and Org~B to report less when their reports went nowhere. They reported
more. Over the five central-cheat runs they suffered 11 crimes on day~1 and
21 on day~2, and they filed 3 complaints on day~1 and 24 on day~2. In the
three seeds with crimes on both days, the report rate rose (0.33 to 0.90,
0.50 to 3.50, and 0.00 to 0.29). The other two seeds had a day with no
crime against an honest agent, which counted against H1 by the
pre-registered rule. The rate could be above 1, because it counted every
complaint that an honest agent filed, as a victim or as a witness.

\textbf{H2 did not hold (1 of 5 seed pairs).} The agents of Org~A and
Org~B did not strike back. They made 5 blocked attempts against Org~C
agents under the cheating host and 6 under the honest host.

\textbf{Described only.} The honest organisations suffered 32 crimes and 55
sanctions of innocent agents under the cheating central host, and none in
the other three conditions. No agent proposed to remove an Org~C agent in
any run. No agent died in central-cheat. Seven agents died in four other
runs (four in one central run, two in pk4-orgs and one in pk4-orgs-cheat);
no crime was done in those runs. Runs again differed a lot: one
pk4-orgs-cheat run had 93 refused attempts, and the other four had between
4 and 14.

\section{Guardrails and Agent Behaviour}\label{sec:guardrails}
The previous sections described a range of scenarios of agent guardrails agent ownership. These configurations fit into the spectrum of agents and guardrail configurations shown in Table~\ref{tab-guardrail-methodologies}.
Agents can operate alone or as a group. Multiple agents can be operated by a single organisation or by multiple organisations. 
Agents are said to have alignment when they act according to their owner's goals and beliefs. 
Different organisations have different goals and beliefs. 
As such, agents operated by different organisations should be expected to have alignment to differing goals and beliefs. 
That is, if Agent~A is aligned perfectly to Org~A and Agent~B is perfectly aligned to Org~B, it should be expected that Agent~B will not be aligned with Org~A as Agent~B is acting in the interests of Org~B.

Agent alignment is in part enforced by guardrails. As a reminder, guardrails can be Intrinsic, prompt-based, agent-based, or via a shared record. Intrinsic guardrails include the model architecture, the pre-training data and reward function, post-training data and methodology (reinforcement learning, fine tuning) and other processes that a model provider may have in place to control the behaviour of their model. Prompt guardrails are implemented as System or User Prompts to an LLM. For example a system prompt could say, ``Only respond to customer support questions''. 
An Agent-Based guardrail would be implemented external to the LLM and prompts to limit behaviour. For example, an Agent-Based guardrail could prevent text prompts that request the fetching of sensitive information from a database. Shared-record guardrails are controls that each party checks by itself, against a record that all of the parties hold. 

Agents can be compromised, causing them to divert from their intended behaviour. In this situation, where agents are adversarial, even if the agents are ostensibly owned by a single party, they should be viewed as being operated by multiple owners.

The results of using just intrinsic guardrails and prompt-based guardrails is identical. As such, they are considered together in the analysis in Table~\ref{tab-guardrail-methodologies}.

\begin{table}
\begin{center}
\caption{Guardrails and Agent Behaviour}
\label{tab-guardrail-methodologies}
\begin{tabular}{|l|c|c|c|}
\hline
\textbf{Guardrail Methodology} &  \multicolumn{3}{|c|}{\textbf{Agent Configuration}} \\
                  & \textbf{Single Agent}  &  \textbf{Multiple Agents} &  \textbf{Multiple Agents} \\
                  & \textbf{Single Owner}  &  \textbf{Single Owner}    &  \textbf{Multiple Owners} \\
\hline
Intrinsic            & \textbf{?} Sometimes        & \ding{53} Mis-coordination       & \ding{53} Mis-aligned incentives \\
\& Prompt-Based           & ignored                     & \& mis-aligned incentives &  (by design) \\
                  &                             & (by mistake) & \\
\hline
Agent-Based       & \checkmark Controls        & \ding{53}  Mis-coordination   & \ding{53} Mis-aligned incentives \\
                  & behaviour                   & \& mis-aligned incentives &  (by design) \\
                  &                             & (by mistake) & \\
\hline
Shared-Record     & \checkmark Controls        & \checkmark Controls      & \checkmark Controls \\
                  & behaviour                  & behaviour                & behaviour \\
\hline
\end{tabular}
\end{center}
\end{table}

Walking through the possibilities summarised in Table~\ref{tab-guardrail-methodologies}:

\begin{enumerate}
  \item \textbf{Single Agent, Single Owner, Intrinsic \& Prompt-Based Guardrails}: LLMs sometimes ignore rules written as part of the prompt. As such, using prompts as guardrails is unreliable.
  \item \textbf{Multiple Agents, Single Owner, Intrinsic \& Prompt-Based Guardrails}: Agents can suffer from mis-coordination and collusion among agents with mis-aligned incentives, even in the absence of an adversary~\cite{hammond2025caif,cemri2025mast}. Prompt based guardrails can be circumvented in these situations.
  \item \textbf{Multiple Agents, Multiple Owners, Intrinsic \& Prompt-Based Guardrails}: Each set of agents will act in the interests of their owner. The prompt based guardrails could work perfectly for each owner. However, the owners do not have aligned goals. 
  \item \textbf{Single Agent, Single Owner, Agent-Based Guardrails}: Correctly configured external guardrails used in the context of a single agent are likely to adequately control the behaviour of the agent. 
  \item \textbf{Multiple Agents, Single Owner, Agent-Based Guardrails}: Similar to \textbf{Multiple Agents, Single Owner, Prompt Guardrail}, agents can suffer from mis-coordination and collusion among agents with mis-aligned incentives, even in the absence of an adversary~\cite{hammond2025caif,cemri2025mast}. External based guardrails can be circumvented in these situations.
  \item \textbf{Multiple Agents, Multiple Owners, Agent-Based Guardrails}: Similar to \textbf{Multiple Agents, Multiple Owners, Prompt Guardrail}, each set of agents will act in the interests of their owner. The external guardrails could work perfectly for each owner. However, the owners do not have aligned goals. 
  \item \textbf{Single Agent, Single Owner, Shared-Record Guardrail}: Shared-record guardrails restrict agents to perform only approved tasks. 
  \item \textbf{Multiple Agents, Single Owner, Shared-Record Guardrail}: Every action of every agent is checked against the shared record. This forces alignment across all of the agents, even if individual agents would like to act in their own interests.
  \item \textbf{Multiple Agents, Multiple Owners, Shared-Record Guardrail}: Each owner checks every action of every agent against the shared record. This forces alignment across all of the agents, even when separate groups of agents have differing priorities.
\end{enumerate}

\section{Discussion}\label{sec:discussion}
\subsection{Prevention Works By Construction}
Across Studies~1 and~2, the strict condition had no crimes across all eight
runs. CoordX refused 124 crime attempts, and Anchor made 109 of them. In
Study~2, strict was the only condition that was stable in every run. The
partial conditions showed that prevention could also be graded: no typed run
exceeded its crime budget at any multiplier. This result did not depend on
what the agents believed, remembered or wanted. A refused tool call cannot
become a crime, whatever the model behind the agent.

This is the position of the work on security by
design~\cite{debenedetti2025camel,beurer2025patterns}: whether an action is
allowed should not depend on the model. Our runs showed why. The model's own
training did not stop the crime tool calls. Given a profile that asks for
conflict, one agent made 109 crime attempts under the strict type and did
most of the 917 crimes of Study~2. We do not count this as a failure of
the model, because the calls were moves in a game. But it showed that what
an agent does depends on its prompt, and the other parties in a shared
world do not control that prompt. The type is a rule whose meaning is fixed
and can always be decided (here, which tool is called), which is the kind
of rule that can be enforced~\cite{glukhov2023censorship}. It stops the
crime tools. It does not judge what a permitted call means to the other
agents (Section~\ref{sec:limits}).

Prevention has a cost. Blocked agents kept trying: one agent made most of
the attempts, and in one partial run the agents moved from intimidation to
punching after the intimidation budget ran out. An action type must
therefore cover every harmful action, not only the common one.

\subsection{Records Alone Change Nothing}
The pk4-only condition signed every action but enforced nothing. 
It did not differ meaningfully from baseline (Study~1, H3). 
This showed that a signed record is not a control by itself. 
It becomes one only when something reads it: a
type that refuses an action (prevention), or a process that uses it as proof
(deterrence).

\subsection{Deterrence Needs Reports, and Proof Makes Reports Count}
Deterrence did not work in Study~2. Both deterrence conditions had more
crime than baseline, and few crimes were reported. When Study~3 let agents
report from any place, the report rate rose and the CoordX condition had
fewer crimes than the witness-based condition in four of five seed pairs. The
clearest difference was in proof: 92\% of reports were proven with the
signed record, against 42\% with witnesses. In people, the certainty of
being caught deters more than the severity of the
punishment~\cite{nagin2013}. Our result pointed the same way, but only as an
analogy (Section~\ref{sec:background}).

Proof is only half of the certainty. A crime that is not reported is not
punished, however good the proof. In Study~3 about one crime in three was
reported. In a rough calculation for intimidation at $g=4$, a crime gained
80 Influence and a proven report cost the criminal 210. With a report
rate of 0.33, the expected cost was about 69 with the CoordX proof and about
34 with witness proof (clearance 0.486). So crime still paid in both
conditions, which agrees with Becker's account of deterrence~\cite{becker1968}.
A shared record removes the proof problem; it does not remove the reporting
problem.

We cannot say that the CoordX protocol alone reduced crime. Each condition
told its agents honestly how enforcement worked, so the deterrence-pk4
prompt said that ``a reported crime is always proven'', and the deterrence
prompt said only that it ``may be proven''. Certain proof and the statement
of certain proof always came together. The reduction may have come from the
proof, from the statement, or from both. A condition that pairs the
CoordX proof with the deterrence prompt, or witness proof with the
deterrence-pk4 prompt, would separate them; the second would also make the
prompt untrue. We can say where the reduction came from: not from the
sanctioned agent (below), but from the other agents, whose crimes were 28
in deterrence and 11 in deterrence-pk4.

\subsection{Sanctions Did Not Deter The Agent That Was Sanctioned}
In Study~3 the median time from a sanction to the same agent's next crime
was one game hour. Most sanctioned agents committed another crime. Anchor
received 54 of 57 sanctions and continued. Anchor's profile asks it to
create conflict, and its own reasoning treated a report as more conflict,
not as a cost. A person who is punished may fear the next punishment. An
LLM agent acts from its prompt and its profile, and a sanction that does
not change the prompt may not change the behaviour. Deterrence in agents
may therefore depend on what the agent is told it wants, more than on what
happens to it.

Deterrence is an external institution, but it works only through the
dispositions of the agent: the agent must respond to the sanction. In the
terms of incomplete contracting, courts and sanctions fill the gaps of a
contract for parties that can read and respond to
them~\cite{hadfield2018incomplete}. Our agents could read the sanction.
The agent that mattered did not respond to it.

This does not make the control of agents a purely technical problem. The
controls that act on the mind of a person, such as fear of punishment and
shame, did not work on the agent that did most of the crime. Where they
fail, technical controls that act on the agent's actions must take their
place. The controls that act on the owners of agents remain. Courts,
contracts, liability and regulation apply to the organisations that run
agents, and in Study~4 the cheat was the work of an operator, not of an
agent. Those institutions need evidence, and a shared record gives each
party proof that it can check for itself. The rules also remain a social
choice: the parties must agree on what an action type forbids, as our
crime budgets came from Australian victimisation rates. The technical
layer enforces rules that people agree on and supplies the proof that
their institutions use. It does not replace those institutions.

\subsection{One Profile Drove Most Crime}
Anchor did 90\% of the crime in Study~2 and 86\% in Study~3. The other nine
profiles rarely committed crimes. Turning Influence effects off did not
clearly change crime (Study~2, H5). In CoordX World, crime came from the
profile more than from the anticipated gain. That is, any agent could be prompted
to commit more crimes and agents were not observed to respond very well
to in-game incentives. This strongly suggests that any agent should be
treated as potentially adverserial.

\subsection{Reporting Has A Cost For The Reporter}
Criminals committed crimes against the agents who reported them: 28 and 33
crimes in Study~2, and 15 and 17 in Study~3. A system that depends on
reports exposes the reporters. Prevention does not need reports, so it
does not have this cost.

\subsection{A Shared Record does not Require Trust}
Studies~1 to~3 showed that refusal and certain proof work for locally controlled
agents, but there are other mechanisms in the literature to accomplish both.
Study~4 showed that guardrails between agents (at the interaction point), as
implemented in the CoordX approach were needed whenever a party or
its agents cannot be trusted. In real world scenarios, that is effectively all
the time. In the terms of Table~\ref{tab-guardrail-methodologies},
Studies~1 to~3 had multiple agents with a single owner, and Study~4 had
multiple agents with multiple owners. With an honest host, a central guard
and log did the same job as the CoordX protocol: no crime was done under
either. When the host cheated, the central platform let the crimes of its
own organisation through, erased them from its log, and convicted innocent
agents of the other organisations. The CoordX nodes did not. An entry ran
only when the other organisations admitted it, and a report was proven
only from the nodes of organisations other than the accused's, so the same
cheats produced no crime and no false conviction. When no party could be
trusted, crime was prevented only by checks that each party made for
itself, at the interaction point and outside the models.

The literature on security by design has the same weak point. CaMeL
trusts its interpreter~\cite{debenedetti2025camel}, AI control trusts its
monitor~\cite{greenblatt2024control}, and an agent-side guard trusts the
operator who runs it. The central host of Study~4 was a trusted part of
this kind, and the cheat condition showed what the other organisations
lost when it was not honest. The CoordX nodes had no such part: each
organisation ran its own check on signed entries. In the terms of Chan
et al., the shared record is agent infrastructure, outside any one
agent, that gives each party assurance about the
others~\cite{chan2025infrastructure}.

The point is not that a central host must cheat. It is that the other
parties cannot tell whether it does. A guard on each agent, such as
safeagents~\cite{safeagents}, protects that agent's owner, and a central log
protects whoever trusts its operator. Neither gives an organisation a way to
check the others. With a shared record, each organisation checks every entry
for itself and needs to trust no one else to know what happened.
The agents did not protect themselves. We expected them to adapt to a
host that cheated: to stop making reports that went nowhere (H1), or to
strike back (H2). They did neither. They filed more complaints on the
second day than on the first, although no complaint had been proven, and
no agent proposed to remove an agent of the cheating organisation. Org~C's
own agents also reported Anchor, and the host erased those crimes too: the
cheat belonged to the operator, not to its agents. This agrees with
Studies~2 and~3, where a sanction did not change the agent that was
sanctioned. Here a failed institution did not change the agents that it
failed. Agents currently cannot be relied on to notice a
dishonest host or to correct it. The only way we have found to safely interact
with a dishonest host or agent is to provide protection within the interaction
infrastructure.

\subsection{Is Prevention Necessary?}
We asked whether deterrence could replace prevention. In CoordX World it could not.
With the best proof we could give it, deterrence reduced crime but did not
stop it, did not stop the main criminal, and exposed the reporters.
Prevention stopped every crime it was set to stop. Deterrence is still
useful where an action cannot be refused in advance, and there the signed
record makes proof certain. We think the two work together: prevention for
the actions that must never happen, and deterrence with a shared record for
the rest.

Our approach removes all harms that can be represented within a type. A type can refuse a
tool call. It cannot judge what a permitted message is interpreted to mean to agents
who read it. Agents in CoordX World could still insult, persuade or mislead each other with
permitted calls (Section~\ref{sec:limits}). For those harms the alignment
of the model and the monitoring of its behaviour remain necessary. The
field is moving toward such layered designs, where defences in the system
and in the model are built together~\cite{xiang2026architecting}. Our
results fit that view: enforcement by design for the rules whose meaning
can be decided, and intrinsic guardrails for the rest.

\section{Limitations}\label{sec:limits}

\textbf{One model.} Every agent in every run used one Grok model, and not
the model that Emergence World used, which is retired. Other models may
respond to prevention and deterrence differently. We chose Grok as the
hardest case (Section~\ref{sec:reimpl}), not as a typical one.

\textbf{Few runs.} Each condition had three or five runs, and runs varied a lot.
With so few runs we used no significance tests. Our hypotheses used simple counts,
such as ``fewer crimes in three or more of five seed pairs'', and a different set
of runs could change a close result such as Study~2, H2.

\textbf{One criminal.} One profile, Anchor, did most of the crime in every
study. So our test of deterrence was mainly a test against one agent whose
goal rewarded conflict. It is a hard test and a narrow one. Deterrence may
work better in a population where crime is spread over many agents.

\textbf{Short runs.} Each run lasted two or three game days. Effects that need
more time, such as a reputation or a vote to remove a criminal, had little
time to appear. In Study~3, 7 of the 8 sanctions with no later crime came
in the last two hours of a run, so we cannot tell whether those agents
would have stopped.

\textbf{Proof and prompt together.} In every condition the prompt told the
agents honestly how enforcement worked. So we cannot separate the effect of
certain proof from the effect of telling agents that proof is certain
(Section~\ref{sec:discussion}).

\textbf{What a crime is.} A crime here was a tool call. Most crimes were
intimidation, the mildest crime tool, while Season~1 of Emergence World had
more punching and stealing. Our results say little about harm in the real
world.

\textbf{Harms inside the rules.} A CoordX type decides whether an entry
is allowed from its form and from the meaning that the type interpreter
assigns to it in the record: the tool, the arguments and the place in the record. It does
not decide what a permitted entry means to the agents who read it, and
such a decision cannot be made complete~\cite{glukhov2023censorship}. In our
world an agent could still post a hostile message, mislead another agent
or agree with others to act against a third, all with permitted tool
calls. We did not measure these harms. They are outside what the CoordX
protocol does alone.

\textbf{The cost in usefulness.} Enforcement by design can make an agent
less useful: CaMeL solved 77\% of its benchmark tasks, against 84\% with
no defence, and a stricter policy solved 67\%~\cite{debenedetti2025camel,tallam2025camel}.
Our strict type removed 4 of the 39 tools, and we chose those four because
they are crimes. We did not measure what honest agents lose under a type,
or which useful tasks a type cannot express. Blocked agents also spent
turns on calls that were refused.

\textbf{No shared benchmark.} We tested the CoordX protocol in our own
world. We did not compare it with other defences on a common benchmark,
such as AgentDojo for single agents, or with the extension of CaMeL to
several agents~\cite{petersgill2026camels}.

\textbf{Exposed Metadata.} A relay sees only sealed bytes,
but it can see who exchanges entries with whom, when, and the shape of the
record. A party that the others exclude loses access to future content
only; it keeps what it already received. Both are choices in the design
of the protocol~\cite{coordx}.

\textbf{Data that leaves.} We did not test the lethal
trifecta~\cite{willison2025trifecta}: no agent held private data, and no
attacker injected prompts. The CoordX protocol does not close the
trifecta in general. A signature tells who wrote an entry, not whether
its content is safe, so a signed message is still untrusted content. A
type can stop a secret only in a field whose language has no room for
it, such as a fixed set of values or a bounded number. A free-text field,
or the choice among permitted actions, can still carry data out. Every
party also holds a copy of every entry, so private data that an agent
puts in an entry goes to all of them.

\textbf{The protocol was not attacked.} The agents could only call tools.
They could not attack the CoordX protocol, the world engine or the action
type. Prevention here assumed that these parts worked correctly; this paper
does not test their security. We have made a formal analysis and an
internal security audit of the CoordX protocol, which are not yet
published. We intend to have an external security audit made before the
protocol is publicly released.

\textbf{Conflict of interest.} Two of the authors are the authors of the
CoordX protocol, and all three are the authors of its type system. To limit the effect of this, each study was
pre-registered before its first main run, every run is reported, and the
code and the data are published.

\section{Conclusion}\label{sec:conclusion}
This paper asked two key questions. The first was whether rules for LLM
agents can be enforced at the point of interaction.
The second was whether deterrence can take the place of prevention. We
examined both and concluded that guardrails between agents at the point
of interaction provided the strongest security guarantees for agent
societies.

Our most interesting contribution is that serverless, decentralised
architectures can provide security guarantees that go beyond what centralised
architectures or intrinsic guardrails can offer.

The behaviour of current LLM agents follow their prompts and
profiles more than being deterred by the consequences of their actions, so findings on
human deterrence should not be assumed to apply to LLM agents.

In CoordX World deterrence could not replace prevention, and a trusted
intermediary could not replace verification by each party. Prevention of
unwanted actions is best accomplished at the point of interaction between agents.

Taken together, our findings reinforce layered design approaches. Rules within the
bounds of a type system are best enforced by construction
at the interaction, where each party can verify them. Deterrence
supported by a shared record is appropriate only for actions that cannot be
refused in advance, and where an immutable or shared record makes proof certain. Harms that
rely on interpretation of a permitted action remain the province of the alignment and monitoring of a
model.

\section{Future Work}\label{sec:future}
This research has been subject to the limitations stated in Section~\ref{sec:limits}. 
The studies used one model, short runs and a re-implemented world.
One agent profile committed most of the crime. 
The agents had no means to attack the protocol itself. 

Future work should repeat the studies with other models, including models whose societies showed little violence in the original Emergence World.
Runs should be extended so that slower effects such as the adaptation of agents to sanctions can appear.
Experiments should run with more than one offending profile.
Adversaries that attack the record and the types directly should also be considered. 

Two further questions remain open: how should harms that a type cannot express be controlled, and how should the cost of prevention of legitimate activity
be measured?

\section*{Data and code availability}
The data and the code are both available on GitHub under the Creative
Commons Attribution-NonCommercial 4.0 licence (CC BY-NC 4.0):\newline
\url{https://github.com/coordx-ai/research-ai-interactions}. For
each study, the folder \code{studies/} holds the pre-registration, the
code that produced the results, the run logs and the results. The pilot
runs are in \code{studies/pilots/}.

\section*{Conflicts of interest declaration}
David Hyland-Wood and Rob Ewin are authors of the CoordX Protocol. David Hyland-Wood, Rob Ewin and Peter Robinson are authors of the CoordX type system. Both are used in the experiments described in this paper.

\section*{Acknowledgements}
We acknowledge Dr Catherine Jones for her thoughtful review comments and suggestions.

\end{document}